\documentclass[aip,reprint]{revtex4-1}
\usepackage{graphicx}
\usepackage{listings}
\usepackage{picins}
\usepackage{latexsym}

\draft 

\begin{document}


\title{Lennard-Jones interatomic potentials for the allotropes of carbon} 



\author{Hui Zhang}
\email[Electronic mail:]{zhope@scut.edu.cn}
\noaffiliation
\author{Zhongwu Liu}
\noaffiliation
\author{Xichun Zhong}
\noaffiliation
\author{Dongling Jiao}
\noaffiliation
\author{Wanqi Qiu}
\noaffiliation
\affiliation{School of Materials Science and Engineering, South China University of Technology, Guangzhou 510640, People's  Republic of China}


\date{\today}

\begin{abstract}
Finding appropriate interatomic potentials which can accurately describe the crystal structure of material is one of important topics in material science. In this paper, several interatomic potentials which comprise of Lennard-Jones (LJ) potentials have been proposed for describing both the crystal structures and the evolution of microstructure of the allotropes of carbon such as diamond and graphite. The validity of these LJ potentials can be checked by molecular dynamics (MD) simulation. For the lattice identification of simulated systems, we have calculated the distribution functions of the angles between one atom and its nearest neighbors and the distances between atoms and checked the atomic arrangements. Our simulated results have clearly demonstrated that we have successfully produced diamond and graphite structures by MD simulations and with the above LJ potentials.
\end{abstract}


\maketitle 

\section{Introduction}
\indent The search for appropriate interatomic potentials which can accurately describe the crystal structure of material and the evolution of its microstructure is one of important topics in condensed physics and material science. With these interatomic potentials and without setting any initial lattice, both the liquid-crystalline phase transition and the crystal structure should be reproduced by computer simulations. However, the search is difficult and the research is still under way. Recent investigation has shown that Lennard-Jones (LJ) solids can show the face centered cubic (fcc) lattice or hexagonal close packed (hcp) lattice and the hcp lattice is more stable than the fcc lattice \cite{Zhang-1}. This leads to one question of whether this LJ potential can be a starting point for constructing complex interatomic potentials for crystalline systems. Let us suppose that we make a mixture of A and B components at a high temperature, and anneal the system to be relaxed for the equilibrium. At this stage, the system is in the liquid state. In our strategy, we don’t set any initial lattice for A and B components and AB systems. The interatomic potentials between A-A, B-B, A-B pairs are only determined by LJ potentials, and the strength of interaction is governed by $\sigma$ and $\epsilon$. We cool the system from the liquid state and anneal the system at every temperature for some time. This treatment is different from the conventional ones and its feasibility can be checked by molecular dynamics (MD) simulation. Here, we have chosen the allotropes of carbon as our targets and it has been expected that their structures can be reproduced by MD simulations. For the lattice identification of simulated systems, we have calculated the distribution functions of the angles between one atom and its nearest neighbors and the distances between atoms and checked the atomic arrangements.\\
 \begin{figure}[h t b p]
\centering
 \includegraphics[width=80mm]{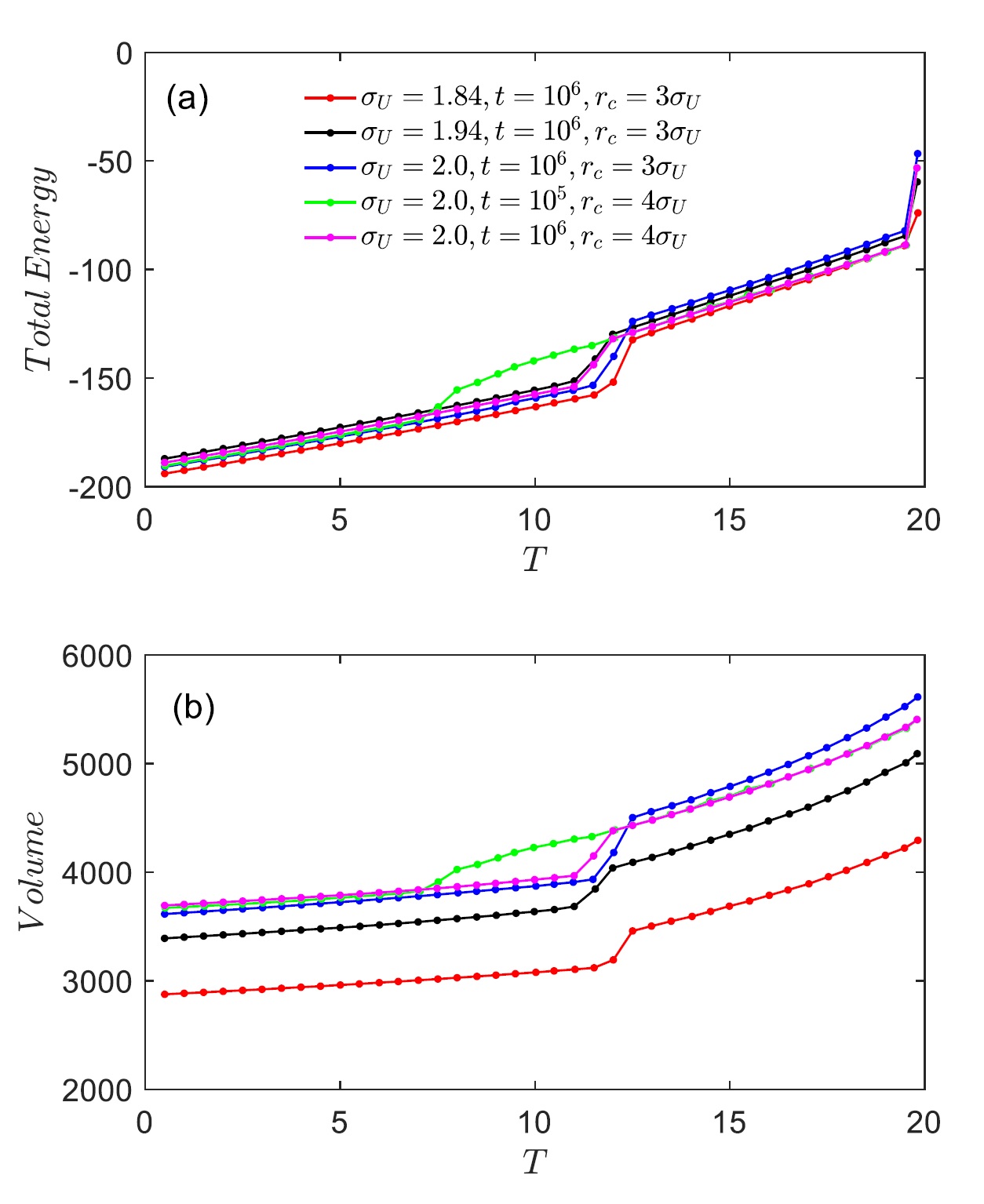}%
 \caption{\label{FIG.1.}Dependence of both the total energy (a) and volume (b) of simulated systems on the temperature for different simulation parameters}
 \end{figure}
   \begin{figure*}
\centering
 \includegraphics[width=160mm]{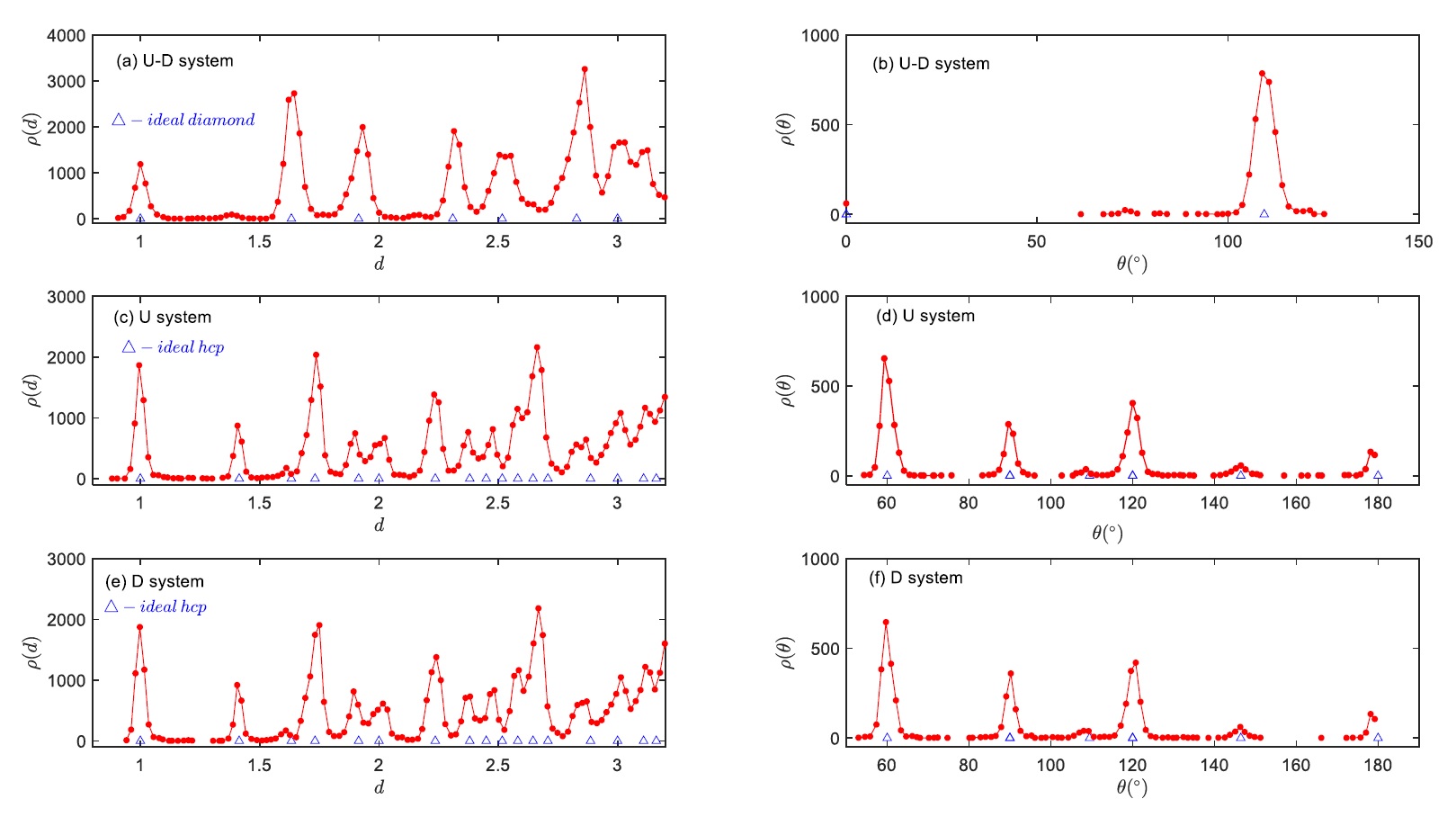}%
 \caption{\label{FIG.2.}Distribution functions of both the angles between one atom and its nearest neighbors and the distances between atoms for $\sigma_{U}$=1.84, $t$=1$\times$10$^{6}$, and $r_{c}$=3$\sigma_{U}$. (a) and (b) are for simulated system, (c) and (d) for U subsystem, and (e) and (f) for D subsystem.}
 \end{figure*}
 \begin{figure*}
\centering
 \includegraphics[width=160mm]{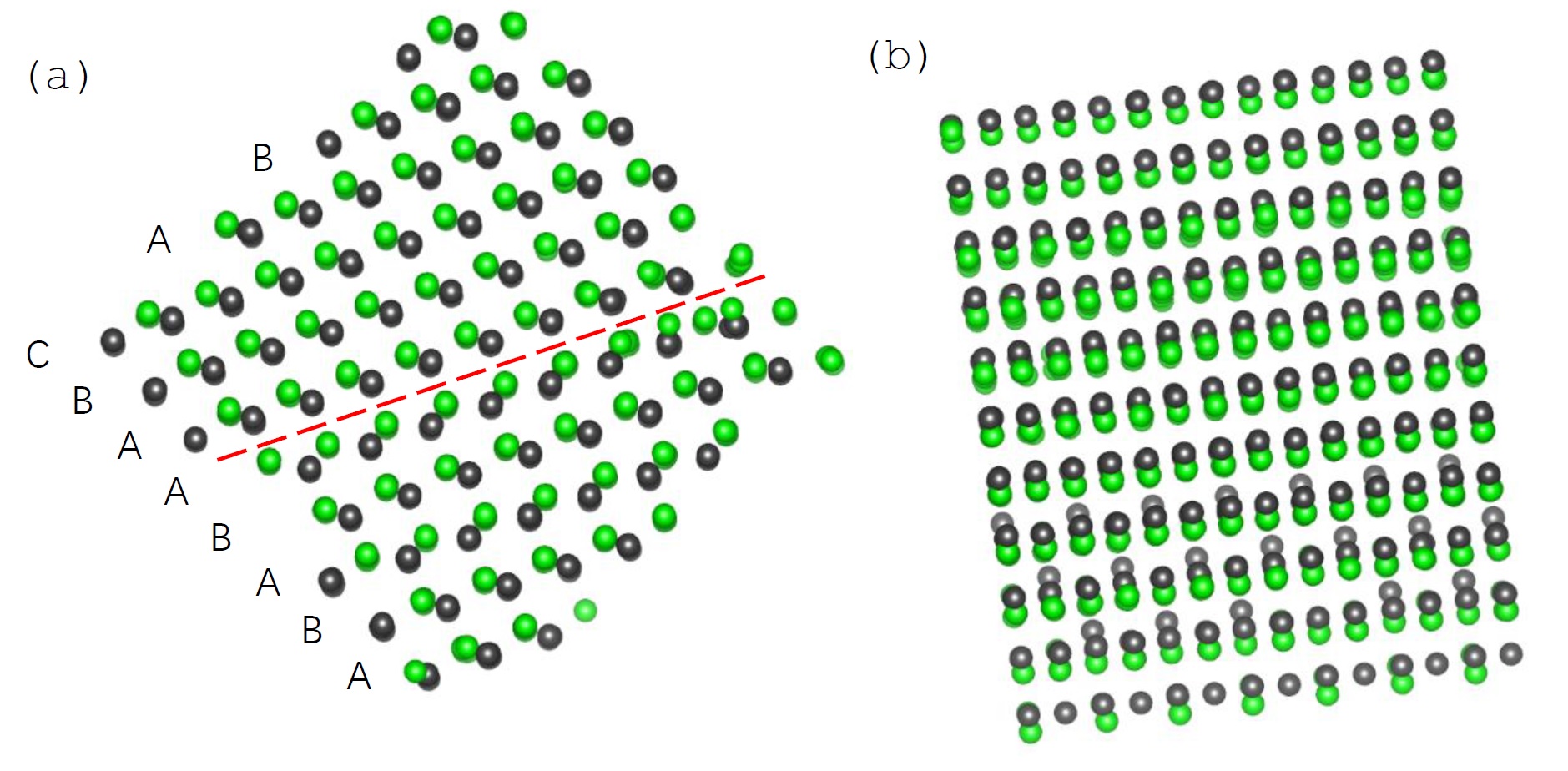}%
 \caption{\label{FIG.3.}The atomic arrangements of simulated system observed from different directions for $\sigma_{U}$=1.84, $t$=1$\times$10$^{6}$, and $r_{c}$=3$\sigma_{U}$. In (a), the atomic arrangements of ABAB$\cdots$ and ABCABC$\cdots$ are presented.}
 \end{figure*}
\begin{figure*}
\centering
 \includegraphics[width=160mm]{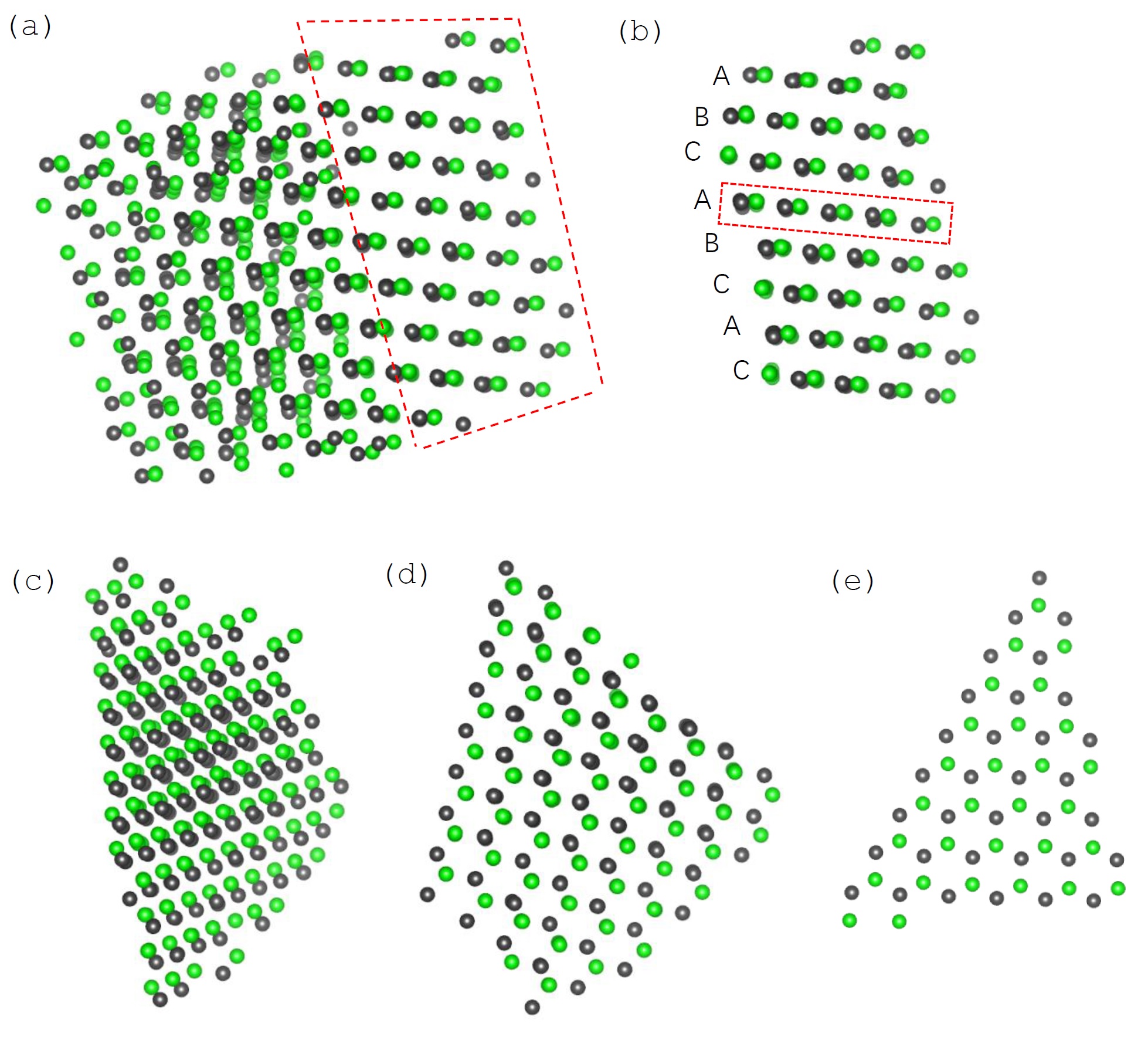}%
 \caption{\label{FIG.4.}The atomic arrangements for $\sigma_{U}$=1.94, $t$=1$\times$10$^{6}$, and $r_{c}$=3$\sigma_{U}$. (a) is for simulated system. (b)-(d) show the atomic arrangements in the region encircle by the red dashed lines in (a) observed from different directions. (e) shows the graphene structure of single layer encircled by the red dashed lines in (b).}
 \end{figure*}
   \begin{figure*}
\centering
 \includegraphics[width=160mm]{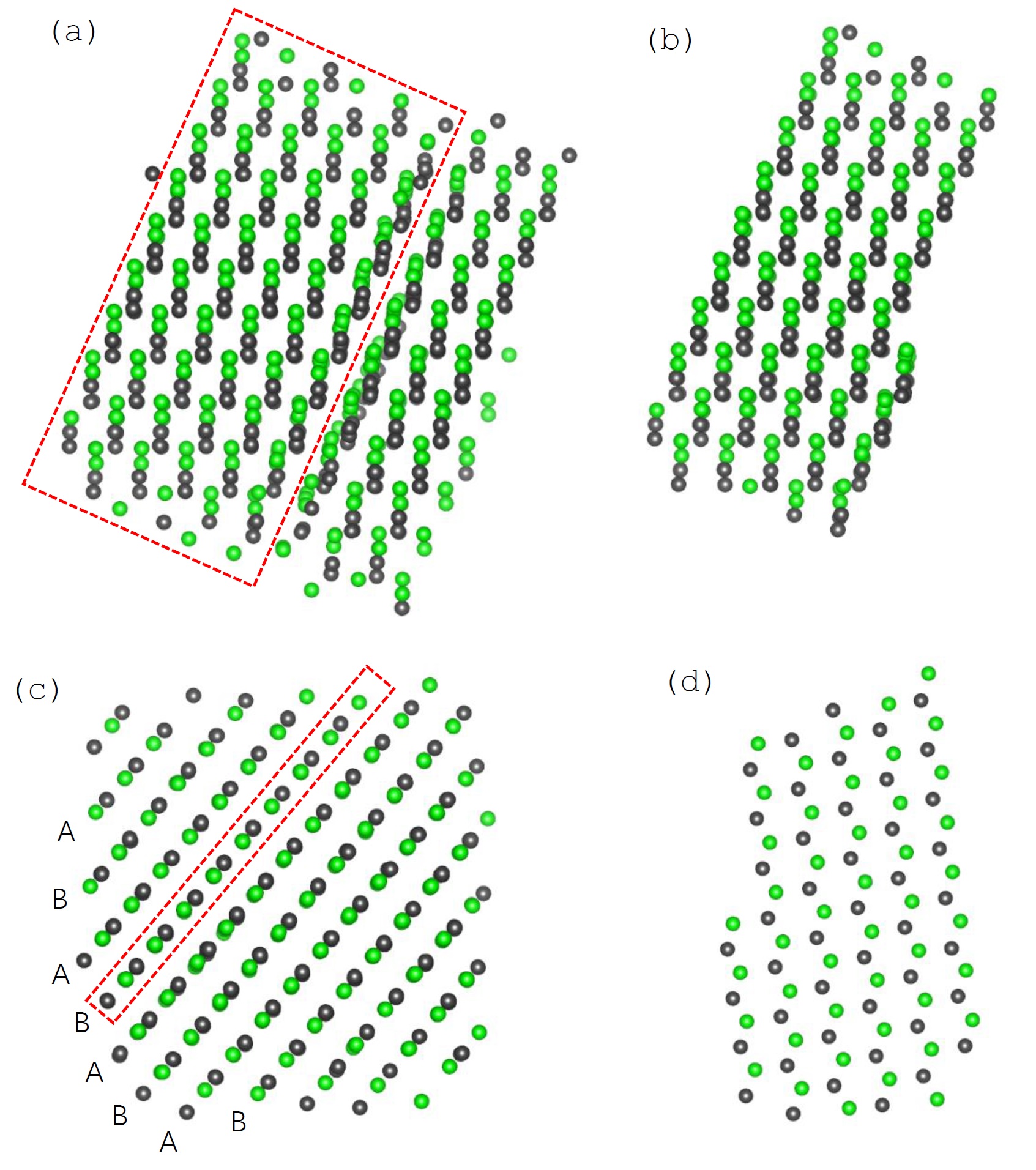}%
 \caption{\label{FIG.5.}The atomic arrangements for $\sigma_{U}$=2.0, $t$=1$\times$10$^{5}$, and $r_{c}$=4$\sigma_{U}$. (a) is for simulated system. (b)-(c) show the atomic arrangement in the region encircled by the red dashed lines in (a) observed from different directions. (d) shows the graphene structure of single layer encircled by the red dashed lines in (c).}
 \end{figure*}
\begin{figure*}
\centering
 \includegraphics[width=160mm]{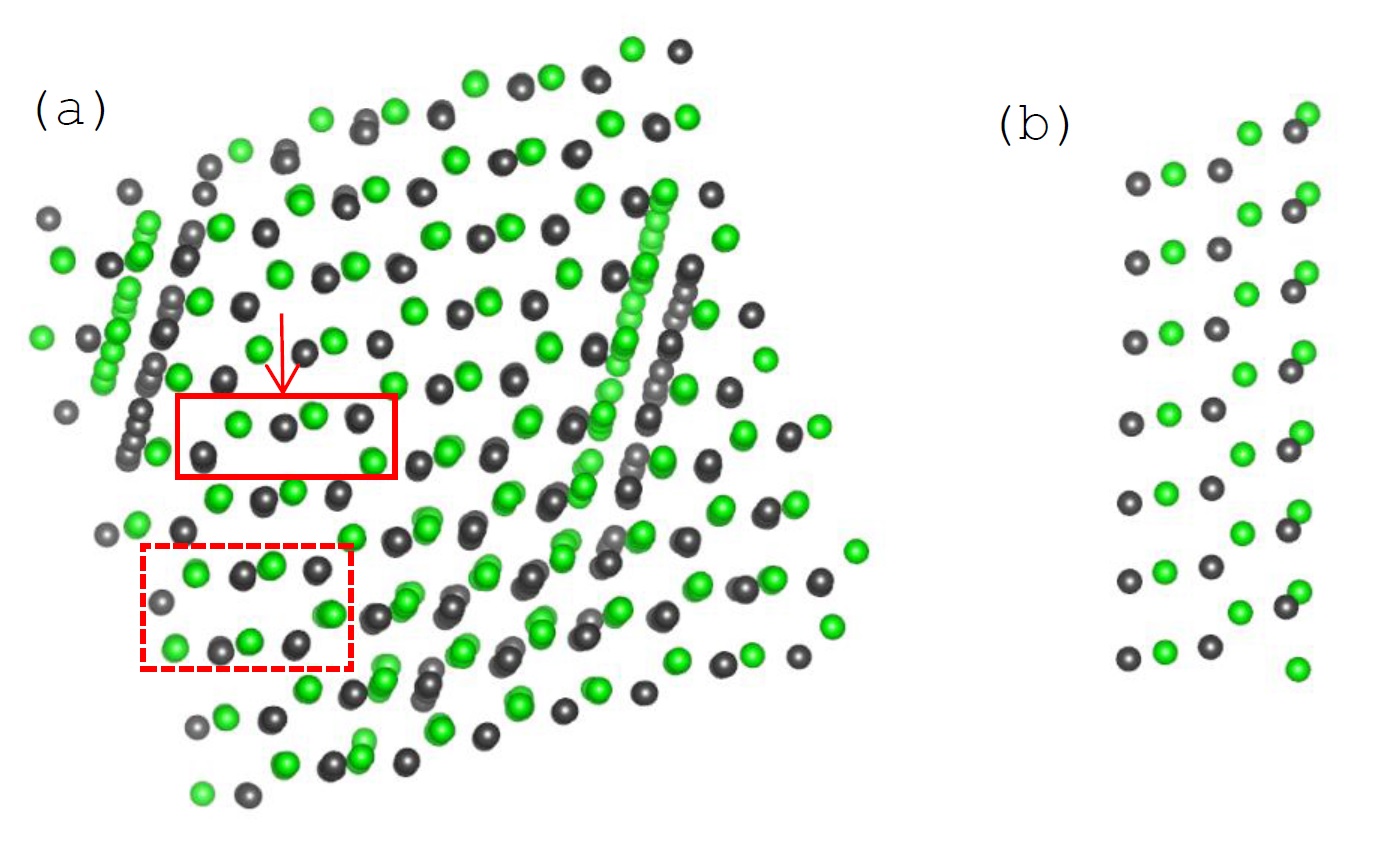}%
 \caption{\label{FIG.6.}(a)the atomic arrangements of simulated system observed from some direction for $\sigma_{U}$=2.0, $t$=1$\times$10$^{6}$, and $r_{c}$=4$\sigma_{U}$. (b)the atomic arrangement in the region encircled by the red dashed lines in (a) observed from the direction indicated by the red solid arrow in (a).}
 \end{figure*}
\begin{figure*}
\centering
 \includegraphics[width=160mm]{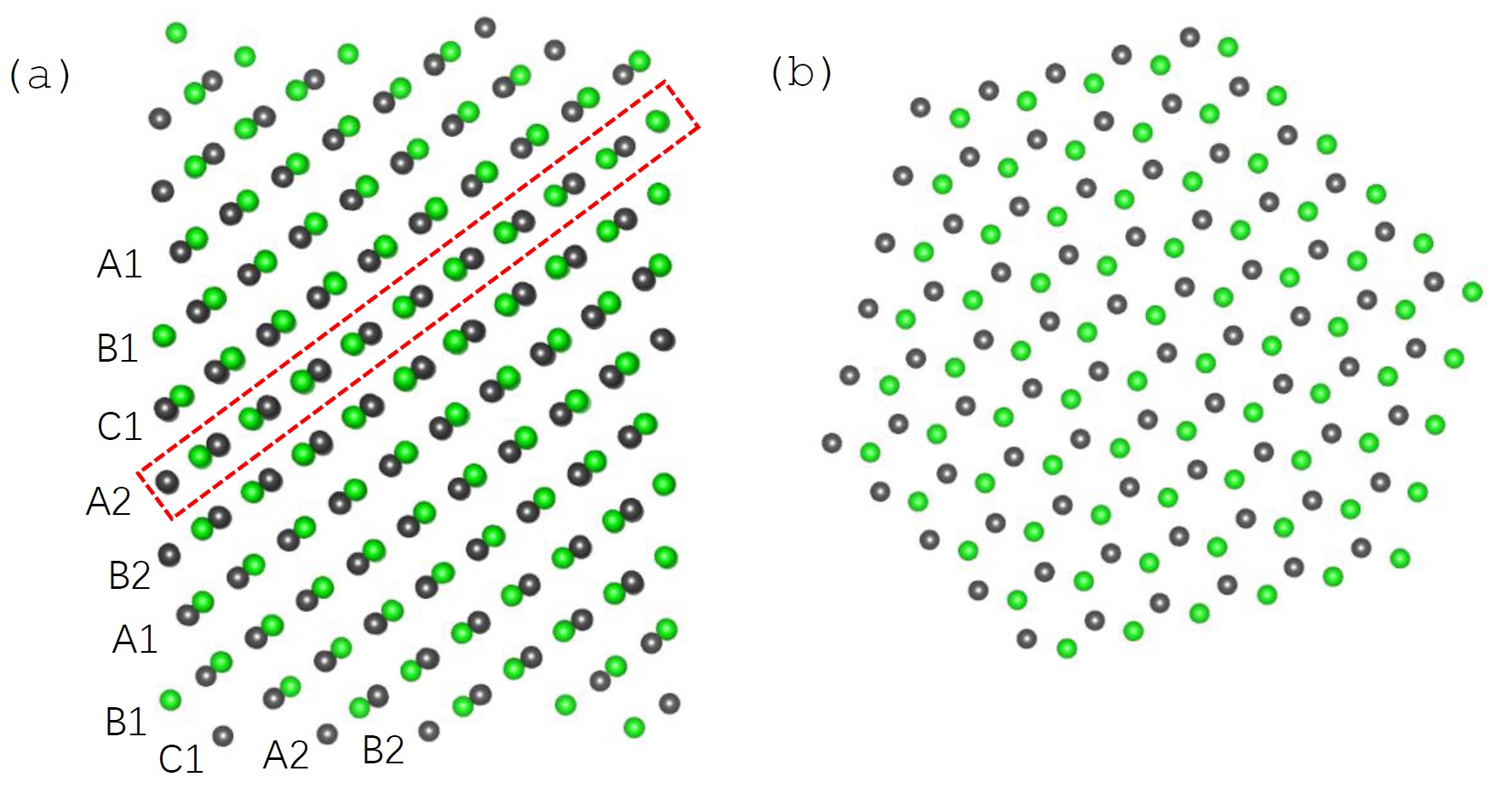}%
 \caption{\label{FIG.7.}(a)the atomic arrangements of simulated system observed from some direction for $\sigma_{U}$=2.0, $t$=1$\times$10$^{6}$, and $r_{c}$=3$\sigma_{U}$. The text in (a) indicates the atomic arrangements of A$_{1}$B$_{1}$C$_{1}$A$_{2}$B$_{2}$A$_{1}$B$_{1}$C$_{1}$A$_{2}$B$_{2}$$\cdots$. Note that the positions of green and black atoms in A$_{1}$, B$_{1}$, C$_{1}$, A$_{2}$, and B$_{2}$ layers are different. (b)the atomic arrangement of single layer encircled by the red dashed lines in (a).}
 \end{figure*} 
 \begin{figure}[h t b p]
\centering
 \includegraphics[width=80mm]{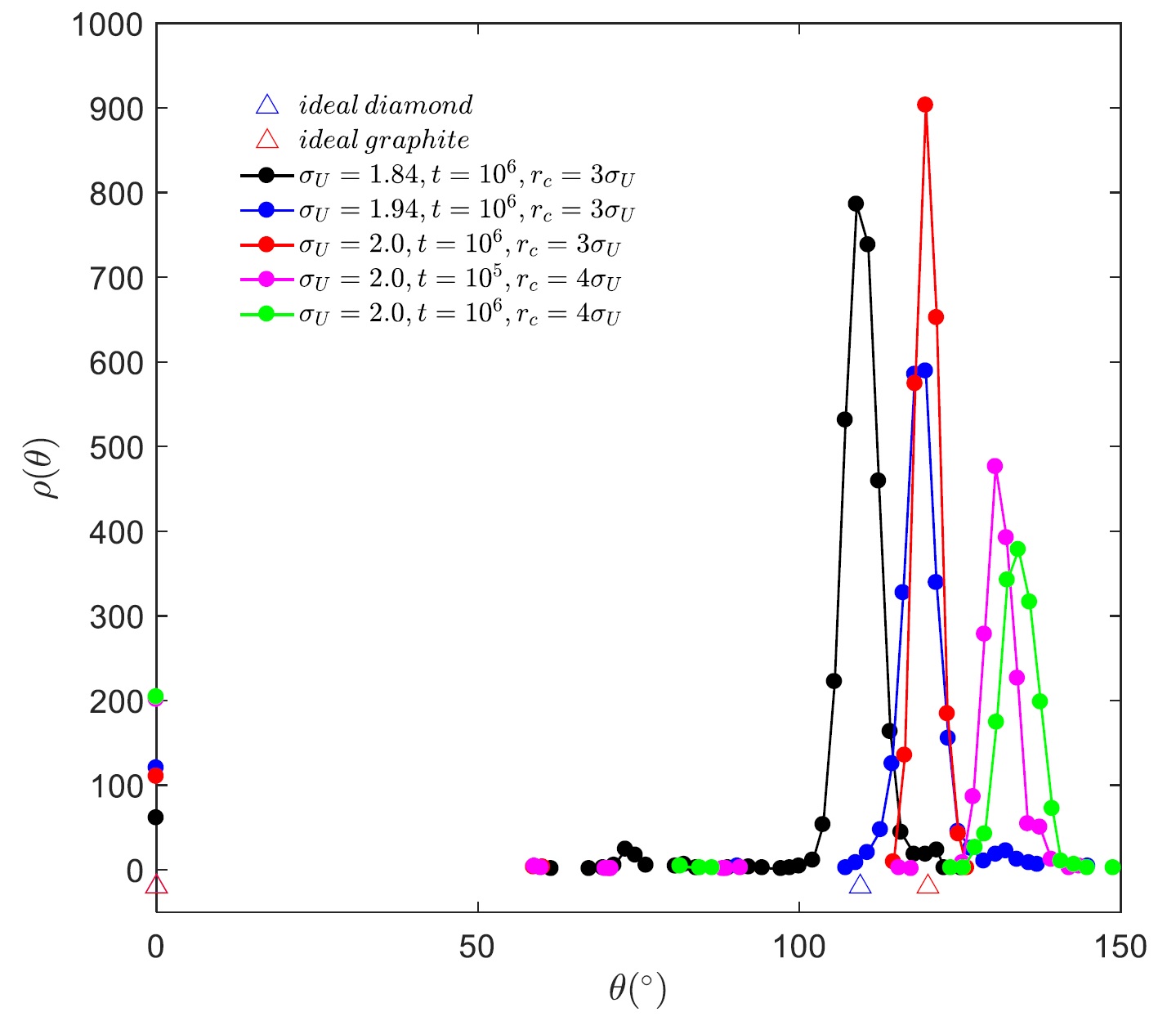}%
 \caption{\label{FIG.8.}Distribution functions of the angles between one atom and its nearest neighbors for simulated systems with different simulation parameters.}
 \end{figure} 
 \begin{figure*}
\centering
 \includegraphics[width=160mm]{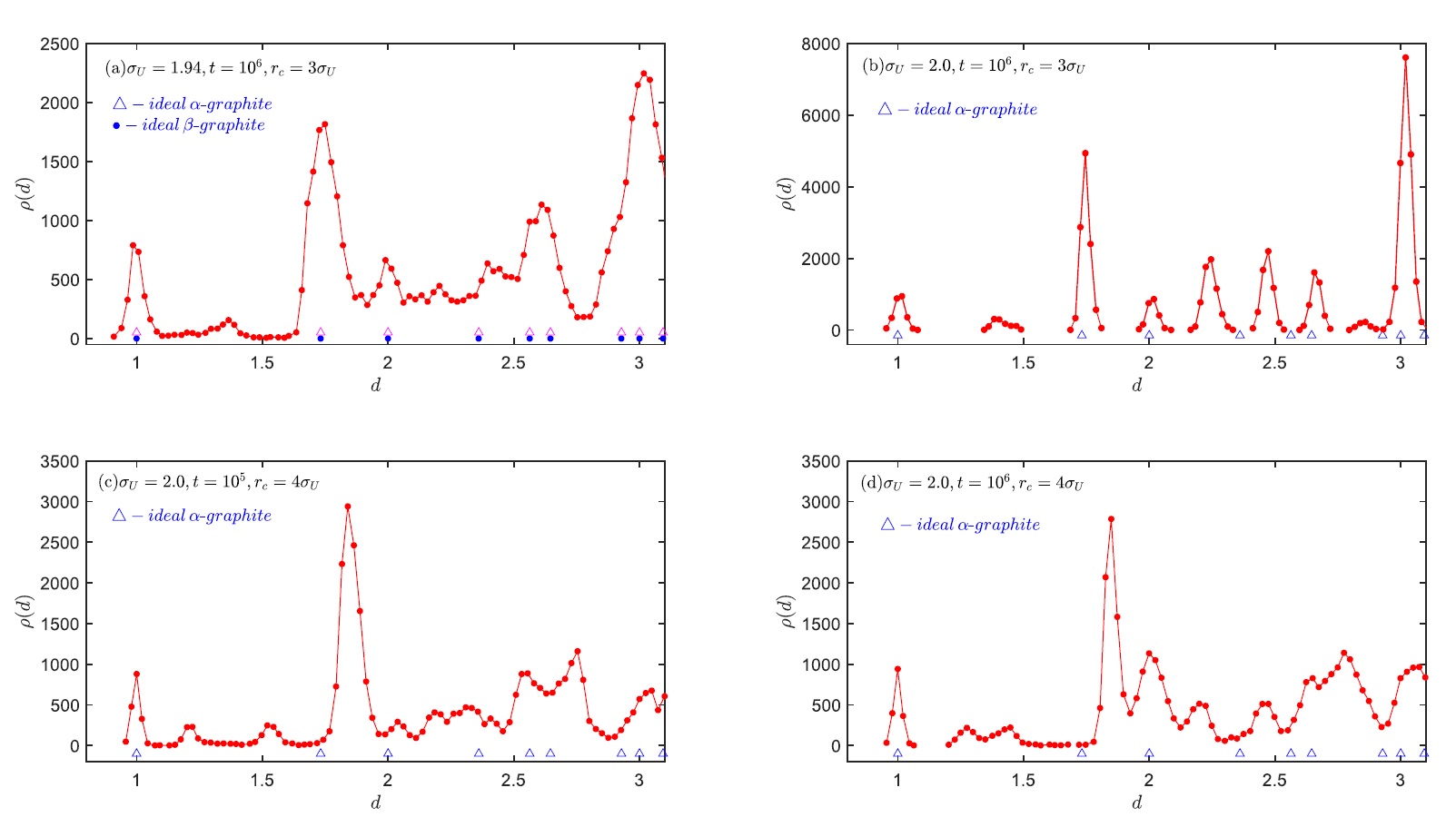}%
 \caption{\label{FIG.9.}Distribution functions of the distances between atoms for simulated systems with different simulation parameters.}
 \end{figure*} 
 
\indent Carbon has two common allotropes. One is diamond, and the other graphite. They show different crystal structures. It is believed that diamond structure is formed under an extreme situation. Graphite shows two crystal structures \cite{Chen-2,Lipson-3,Gupta-4}. One is called the hexagonal $\alpha$-graphite, in which the arrangement of carbon atomic layer is described as ABABAB$\cdots$. The other is rhombohedral $\beta$-graphite with the arrangement of ABCABCABC$\cdots$. Here A, B, and C are single layer of carbon atoms arranged in a hexagonal lattice, and a single layer is also called graphene. Under some circumstances, other allotropes of carbon such as fullerene \cite{Gupta-4,Kroto-5}, carbon nano-tubes \cite{Iijima-6}, and graphene \cite{Novoselov-7} are produced. Spherical fullerene contains both pentagonal and hexagonal carbon rings, and cylindrical fullerene is referred to as carbon nano-tube. These new allotropes of carbon have shown many unusual physical and chemical properties, and are one of the current hot topics in the fields of condensed physics and material science. Their special structures also have attracted our attention. In this paper, our goal is to produce the diamond, graphene, $\alpha$-graphite, $\beta$-graphite, and carbon nano-tube structures by MD simulations and without setting any initial lattice and with only sets of LJ potentials.\\

\section{Simulation and method}
In the simulation, two types of atoms which are physically identical to each other have been introduced in, and named U and D atoms. There is an interaction between U and D atoms. We introduce the classic LJ potential to describe the interatomic coupling, and LJ potential can be written as
\begin{eqnarray}
U(r)=4\epsilon \left(\left(\frac{\sigma}{r}\right)^{12}-\left(\frac{\sigma}{r}\right)^{6}\right)
\end{eqnarray}
where $\epsilon$ is the depth of the potential well, $\sigma$ is the finite distance at which the inter-particle potential is zero, and $r$ is the distance between particles. The distance of cutoff is denoted by $r_{c}$. The simulation was carried out with the aid of LAMMPS \cite{Plimpton-8}. In the simulation, the LJ units were used, and the periodic boundary conditions were applied. $\sigma_{U}$=$\sigma_{D}$=1.84, 1.94, and 2.0. $r_{c}$=3$\sigma_{U}$, and $r_{c}$=4$\sigma_{U}$. $\sigma_{UD}$=1.0 and $r_{c}$=3$\sigma_{UD}$. The masses of the particles were both 100, and the numbers of particles were both 500. We did not set any initial Bravais lattice. The particles were created randomly in the simulation box and then an energy minimization procedure followed. For every value of $\epsilon$, the initial temperature $T_{0}$=$\epsilon$, and $\epsilon$=20. We set the timestep as 0.001. At $T_{0}$, NPT dynamics was implemented for 1$\times$10$^{6}$ or 1$\times$10$^{7}$ timesteps, and then the temperature was decreased by $T_{0}$/40. At every following temperatures $T$, NPT was carried out for a time of $t$ timesteps, and $t$=10$^{5}$, 10$^{6}$, and 10$^{7}$. The pressure was always zero in the simulation. Details can be found in in-script in Appendix. The visualization of simulated results was done with the aid of VESTA \cite{Momma-9}. \\
\indent We calculated the distribution functions of both the angles between one atom and its nearest neighbors and the distances between atoms for the identification of microstructure of simulated system \cite{Zhang-1}. For more information, a set of U (or D) atoms was treated as a subsystem, and called U (or D) system.  Simulated system was a union of both U subsystem and D subsystem, and called U-D system. For simulated system, U and D atoms were treated as the same atoms.
\section{Results and Discussions}
Figure 1 shows the dependence of both the total energy and volume of simulated systems on the temperature for different simulation parameters. As shown in Fig. 1, both the total energy and volume of simulated systems decrease with decreasing the temperature. The liquid-solid phase transition occurs at the crystallization temperature. Also in Fig. 1, the volumes of simulated systems change with the variation of $\sigma_{U}$. The volumes of simulated systems increase with increasing the values for $\sigma_{U}$. Figure 2 shows the distribution functions of both the angles between one atom and its nearest neighbors and the distances between atoms for $\sigma_{U}$=1.84, $t$=1$\times$10$^{6}$, and $r_{c}$=3$\sigma_{U}$. In Figs. 2(a) and (b), both U and D atoms are seen as the same atoms. For simulated system, the angles between one atom and its nearest neighbors are 109.5$^{\circ}$, which is in agreement with those of diamond structure. As in Fig.2 (a), the ratios of the distances between atoms are also in agreement with the data for diamond structure. Our initial identification is that simulated system shows diamond structure. For U (or D) subsystem, it has been shown in Fig. 2 that its crystal structure may be either fcc or hcp lattice for they show similar distribution functions \cite{Zhang-7}. We can check the atomic arrangements in Fig. 2 for the final identification. Figure 3 shows the atomic arrangements of simulated system observed from different directions for $\sigma_{U}$=1.84, $t$=1$\times$10$^{6}$, and $r_{c}$=3$\sigma_{U}$. As shown in Fig. 3(a), this is the typical characteristic of diamond structure. In fact, every U atom has four D atoms as its nearest neighbors, and they together form a tetrahedron with U atom in its body center and four D atoms at its four vortices. As shown Fig. 3(a), both abcabc$\cdots$ and abab$\cdots$ atomic arrangements can lead to the diamond structure, where a, b, and c are close packed atomic layers. The region above the red dashed line shows fcc lattice while the region below is hcp lattice.\\
\indent In diamond structure, each carbon atom has four nearest neighbors, and they together form a tetrahedron. In graphite, each carbon has three nearest neighbors, and the angles between one atom and its nearest atoms are 120$^{\circ}$. A graphene is a single layer of carbon atoms arranged in a hexagonal lattice. The stacking order of graphenes is ABAB$\cdots$ for $\alpha$-graphite and ABCABC$\cdots$ for $\beta$-graphite. As in Fig. 4(b), the atomic arrangements of ABCABC$\cdots$ABC are present, and each layer is a graphene, as shown in Fig. 4(e). When observed from some direction, there are network structures formed by hexagonal tubes which are the typical characteristic of $\beta$-graphite. Such a structure cannot be observed in $\alpha$-graphite. In Fig. 5(c), it has been seen that the atomic arrangements of ABAB$\cdots$ are present, and each layer is similar to a graphene. However, in this structure, the angles between one U atom and its nearest D atom neighbors are not 120$^{\circ}$. One angle is larger than 120$^{\circ}$, and the other two are smaller than 120$^{\circ}$, as shown in Fig. 5(d). Large angles mean small distances between atoms. This structure shows an atomic arrangement similar to that of $\alpha$-graphite. Fig. 5(b) shows one typical characteristic of $\alpha$-graphite, which helps us to identify crystal structure of $\alpha$-graphite. As in both Figs. 4 and 5, it has been found that the lattices of both U and D systems play a big role in forming $\alpha$-graphite and $\beta$-graphite. In $\beta$-graphite, their lattices are both fcc lattice (in Fig. 4(b)), and in $\alpha$-graphite, their lattices are hcp lattice (in Fig. 5(c)). \\
\indent In the simulation, the microstructure of graphite is strongly dependent on the simulation parameters and both the annealing time $t$ and the distance of cutoff $r_{c}$ have significant effects on the graphite. Some metastable structures occur for $\sigma_{U}$=2.0, $t$=1$\times$10$^{6}$, and $r_{c}$=4$\sigma_{U}$. In the region encircled by the red dashed lines shown in Fig. 6(a), there are hollow tubes formed by the intersection of two hexagonal tubes. In such a structure, carbon atoms will not be on the same plane, and the angles between one atom and its nearest neighbors are greater than 120$^{\circ}$ (see Fig. 8). This structure is similar to that of carbon nano-tubes obtained experimentally. There is a significant change in microstructure of simulated system for $\sigma_{U}$=2.0, $t$=1$\times$10$^{6}$, and $r_{c}$=3$\sigma_{U}$. In Fig. 7, each layer is graphene. However, in the A$_{1}$, B$_{1}$, and C$_{1}$ layers, green atoms are on the right and black atoms on the left. In the A$_{2}$ and B$_{2}$ layers, their positions change. The next layer starts from the A$_{1}$ layer, and the atomic arrangement is described as A$_{1}$B$_{1}$C$_{1}$A$_{2}$B$_{2}$A$_{1}$B$_{1}$C$_{1}$A$_{2}$B$_{2}$$\cdots$. Such a structure is more complex and is very different from those of $\alpha$-graphite and $\beta$-graphite. \\
\indent From the above results, the angles between one atom and its nearest neighbors increase and the ratios of the second distance $d_{2}$ and the first distance $d_{1}$ between atoms also increase with increasing $\sigma_{U}$, as shown in Figs. 8 and 9. In Fig. 9, due to the difficulties in obtaining the according ideal structures, the data from ideal $\alpha$-graphite structure are provided for a comparison. It has been shown in Fig. 9 that there are two small peaks occurring between the first distance $d_{1}$ and the second distance $d_{2}$. This may be caused by the wrong atomic arrangement.\\
\indent It has been pointed out that we have introduced in two types of atoms called U and D, and they are physically identical to each other. For these atoms, both electric charge and electric spin are not involved, and the interatomic potentials used are spherically symmetrical. Therefore, it is very surprising that both complex diamond and graphite structures can be produced by MD simulation with these simpler potentials, and the underlying mechanism needs to be further clarified.\\
\section{Conclusions}
Without setting any initial Bravais lattice and with the simpler LJ potentials, we successfully produced the diamond, graphite, and graphene structures by MD simulations. 
%
%
%
\appendix
\section{in script}
\begin{lstlisting}
units         lj
boundary      p p p
atom_style    atomic
dimension     3
region        box block 0 10 0 10 0 10
create_box    2 box
create_atoms  1 random 500 245 box
create_atoms  2 random 500 285 box
timestep      0.001
thermo        1000
group         big1 type 1
group         big2 type 2
mass          1 100
mass          2 100
variable      lcut1 equal 3.0
variable      lcut2 equal 5.52
pair_style    lj/cut ${lcut2}
pair_coeff    1 1 20 1.84 ${lcut2}
pair_coeff    2 2 20 1.84 ${lcut2}
pair_coeff    1 2 20 1.0 ${lcut1}
minimize      1.0e-10 1.0e-10 1000000 &
              1000000
dump          1 all image 1000 image.*.&
              jpg type type zoom 1.6
run           1000
undump        1

variable      ltemp equal 20
velocity      all create ${ltemp} 314029 &
              loop geom
fix           1 all npt temp ${ltemp}  &
              ${ltemp} 1.0 iso 0.0 0.0 1.0
dump          1 all image 500000 image.*&
              .jpg type type zoom 1.6
dump          2 all xyz 500000 file1.*.xyz
run           1000000
undump        1
undump        2

variable      lpa equal ${ltemp}
label         loopa
variable      a loop 39
variable      tem equal ${lpa}-0.50*$a
fix           1 all npt temp ${tem} &
              ${tem} 1.0 iso 0.0 0.0 1.0
dump          1 all image 50000 image.*&
              .jpg type type zoom 1.6
dump          2 all xyz 50000 file1.*.xyz
run           1000000
undump        1
undump        2
next          a
jump          SELF loopa
\end{lstlisting}
\begin{acknowledgments}
This work is supported by the National Natural Science Foundation of China (Grant No. 11204087).
\end{acknowledgments}

\end{document}